\documentclass[conference,anonymous=true,review]{IEEEtran}
\IEEEoverridecommandlockouts
\usepackage{cite}
\usepackage{amsmath,amssymb,amsfonts}
\usepackage{algorithmic}
\usepackage{graphicx}
\usepackage{textcomp}
\def\BibTeX{{\rm B\kern-.05em{\sc i\kern-.025em b}\kern-.08em
    T\kern-.1667em\lower.7ex\hbox{E}\kern-.125emX}}
\usepackage{subcaption}
\usepackage{multirow}
\usepackage{xspace}
\usepackage{url}
\usepackage{xcolor}
\usepackage{hyperref}
\usepackage{fancyhdr}
\usepackage{balance}
\usepackage{booktabs}
\usepackage{enumitem}
\usepackage{amsthm, amssymb}
\theoremstyle{definition}
\newtheorem{definition}{Definition}[section]

\newcommand{\modelname}{\textsf{IGRec}\xspace}
\fancypagestyle{FirstPage}{

\lfoot{979-8-3503-2445-7/23/\$31.00 \copyright2023 IEEE}

}
\begin{document}

\title{Group-Aware Interest Disentangled Dual-Training for Personalized Recommendation}

\author{
    \IEEEauthorblockN{Xiaolong Liu$^{1}$, Liangwei Yang$^{1}$, Zhiwei Liu$^{2}$, Xiaohan Li$^{3}$, Mingdai Yang$^{1}$, Chen Wang$^{1}$, Philip S. Yu$^{1}$}
    \IEEEauthorblockA{$^{1}$University of Illinois Chicago, Chicago, IL, USA
    \\\{xliu262, lyang84, myang72, cwang266, psyu\}@uic.edu}
    \IEEEauthorblockA{$^{2}$Salesforce AI Research,
    Palo Alto, USA
    \\zhiweiliu@salesforce.com}
    \IEEEauthorblockA{$^{3}$Walmart Global Tech, Sunnyvale, CA, USA
    \\xiaohan.li@walmart.com}
}

\maketitle

\begin{abstract}
Personalized recommender systems aim to predict users' preferences for items. It has become an indispensable part of online services. Online social platforms enable users to form groups based on their common interests. The users' group participation on social platforms reveals their interests and can be utilized as side information to mitigate the data sparsity and cold-start problem in recommender systems. Users join different groups out of different interests. In this paper, we generate group representation from the user's interests and propose \textbf{\modelname} (\textbf{I}nterest-based \textbf{G}roup enhanced \textbf{Rec}ommendation) to utilize the group information accurately. It consists of four modules. (1) Interest disentangler via self-gating that disentangles users' interests from their initial embedding representation. (2) Interest aggregator that generates the interest-based group representation by Gumbel-Softmax aggregation on the group members' interests. (3) Interest-based group aggregation that fuses user's representation with the participated group representation. (4) A dual-trained rating prediction module to utilize both user-item and group-item interactions. We conduct extensive experiments on three publicly available datasets. Results show \modelname can effectively alleviate the data sparsity problem and enhance the recommender system with interest-based group representation. Experiments on the group recommendation task further show the informativeness of interest-based group representation.
\end{abstract}


\section{Introduction}
\thispagestyle{FirstPage}
Recommender systems (RS)~\cite{rec1,rec2,rec3} are becoming indispensable to web applications owing to their prominent ability in user retention~\cite{DBLP:reference/rsh/ShaniG11} and commercial conversion~\cite{DBLP:journals/tmis/Gomez-UribeH16}. Data sparsity and cold-start problems~\cite{guo2013integrating,cold3,DropoutNet} are still obstacles that most RS suffer from. Data sparsity indicates users can only consume an extremely small proportion of all items on the platform. The cold-start problem with long-tail and incoming users is more severe for the bare information we can acquire. 
Presently, several social platforms allow users to join groups out of their shared interests in items.
Users' group participation is a direct signal to reflect their interests in items, which is informative side information to alleviate the above two obstacles~\cite{WangFGCH19,yin2020overcoming,AGREE} in RS.

\begin{figure}[htbp]
    \centering
    \includegraphics[scale=0.3]{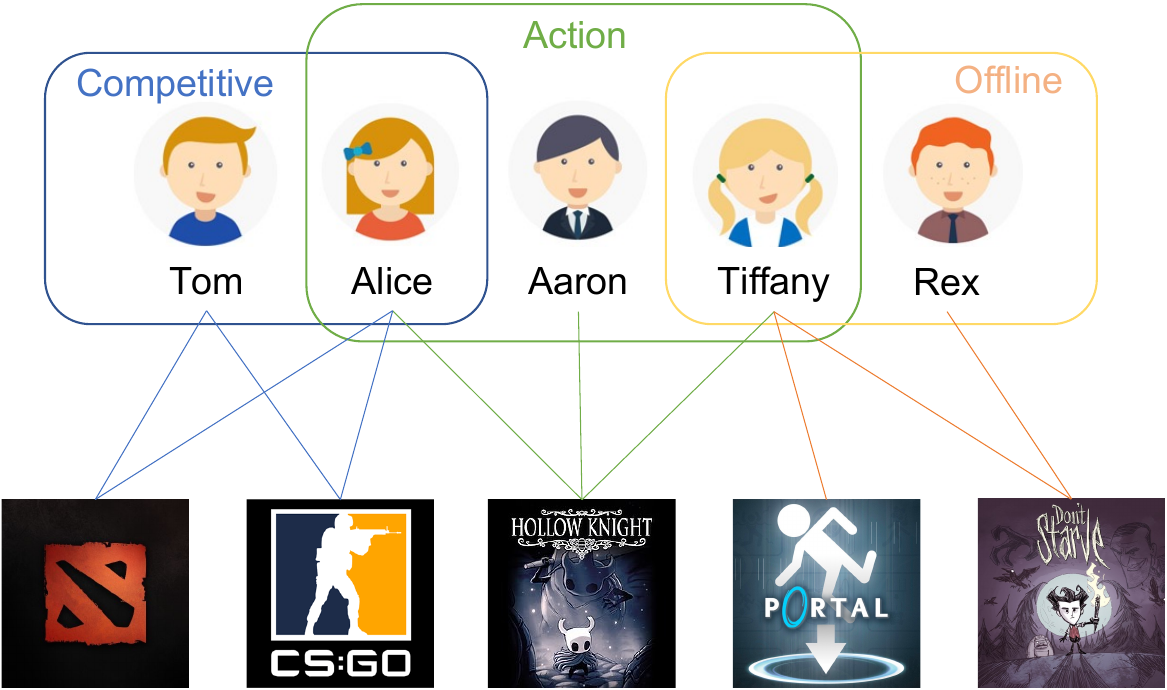}
    \caption{A toy example of the groups of users on Steam Platform. In this example, $5$ users play $5$ kinds of games. Users form "Competitive", "Action" and "Offline" groups based on their gaming preferences.}
    \label{fig:mesh1}
\end{figure}

Personalized RS~\cite{personalized1,personalized2,personalized3,chen2021temporal,li2021hyperbolic} aims to recommend unique items for each user based on his/her behaviors and profile.
The group participation of users on social platforms is personalized behavior that can reflect his/her interests in items.
An illustrative example from the Steam platform~\footnote{\url{https://store.steampowered.com/}} is shown in Figure~\ref{fig:mesh1}: Alice, Aaron, and Tiffany are keen on the action game \textit{Hollow Knight}, and therefore they form a group to discuss the game. It explicitly reflects members' interest in action games.
A user can join multiple groups for different interests in items. For example, Tiffany not only plays the action game \textit{Hollow Knight} but also engages in two other offline games \textit{Portal} and \textit{Don't Starve}. On those platforms, groups can also host activities for group members based on common interests and have direct interaction with items. In Figure~\ref{fig:mesh1}, competitive groups can host online meetups for the ``Dota2'' game competition.
The availability of users' group participation and group activities motivates industrial and academic researchers to utilize it as side information to enrich user embedding~\cite{yuan2009augmenting, DBLP:journals/ipm/LeeB17, GGRM,group1}. These models assume users who participate in the same group share the same preferences. Hence, the user representations are not only learned from interacted items but also from other group members.

Though utilizing group information in RS has attracted academic attention, several key challenges are less explored in effectively exploiting group information: \textbf{(1) Multiple user interests.} Users join multiple groups out of different interests, which are latent and entangled in initial user embedding. For example, in Figure~\ref{fig:mesh1}, Alice joins the `competitive' group because she likes competing with others online in the game CS:GO, while she also joins the `action' group since she enjoys controlling action games. Disentangling and characterizing users' interests in joining different groups enables a more accurate interest-based group representation.
\textbf{(2) Heterogeneous group members.} Previous works~\cite{GGRM,group1} directly aggregate the embedding of its members into group representation, which intrinsically assumes that members in the same group share homogeneous interests. However, it is far from real-world scenarios.
Users in the same group only indicate their interests are partially overlapped. 
Ignoring the heterogeneity of group members within the group would misconstrue the information of this group, thus greatly impairing the quality of a group representation.  
\textbf{(3) Effective group representation for personalized RS.}
The availability of group activities motivates previous research~\cite{GGRM,group1, group_1, group_2, group_3} to focus on recommending items to a whole group, \textit{i.e.}, recommending the same item to a whole group of users rather than personalized RS. How to effectively utilize group activities for personalized RS is still under-explored.

In this paper, we propose a novel framework \modelname  (\textbf{I}nterest-based \textbf{G}roup enhanced \textbf{Rec}ommendation) to tackle the aforementioned research problems. 
\modelname is designed to enhance personalized RS from users' group participation behavior and group activities. As Graph Neural Networks (GNNs) have achieved wide success in RS~\cite{li2020dynamic, liu2020basket, li2021pre, li2022time, li2023impression, DGRec, GraphAU},
the \modelname is built upon a GNN to learn node embeddings over a multi-interest heterogeneous graph that contains user, item, interest, and group nodes.
\modelname consists of four modules.
\textbf{(1) Interest Disentangler} ingests user embeddings effectively for characterizing the crucial interests of users.
It identifies users' interests in joining different groups to address the first challenge of multiple user interests. 
Specifically, we employ a self-gating unit for each type of latent interest, which is trained to extract users' interests from initial embedding.
\textbf{(2) Interest Aggregator} infers the embeddings of those interests aggregated in groups.
This interest aggregator passes those disentangled interests of users and learns interest-based group representation from all group members via a novel Gumbel-Softmax aggregation to address the second challenge of heterogeneous group members on group representation.
Then, since \modelname is built upon the GNN backbone, we stack multiple layers to update the user embeddings by aggregating those interest-based group embeddings from their engaged groups. 
In analogy, the item embeddings are also updated by aggregating all connected users' embeddings. 
\textbf{(3) Interest-based group aggregation} addresses the third challenge on utilizing group representation to enhance user representation. 
\textbf{(4) Dual-trained rating prediction} module unifies the training of both user recommendation and group recommendation tasks. The two tasks train the shared interest disentangler and interest aggregator modules. The dual-trained \modelname can achieve state-of-the-art performance on both personalized and group recommendation tasks. \modelname is open-sourced at \textcolor{blue}{\url{https://github.com/Xiaolong-Liu-bdsc/IGRec}}. In summary, the key contributions of this paper are as follows:
\begin{itemize}[leftmargin=*]
    \item This is the first work that characterizes users' interests in items via group participation, identifies the challenges and proposes the solution of interest-based group representation. 
    \item We propose the interest disentangler via self-gating modules to retrieve multiple interests from user embeddings and a novel interest aggregator to infer interest embeddings.
    \item We devise \modelname based on interest disentanglement. It achieves state-of-the-art performance on both personalized and group recommendation tasks.
    \item To verify the effectiveness, we conduct extensive experiments on two publicly available benchmark datasets MaFengWo and Douban, and extract and publish a ready-to-use group recommendation dataset containing real-world group activity created from the Steam platform~\footnote{To be released upon acceptance}.
\end{itemize}


In the following sections, we first present preliminaries in Section~\ref{sec:pre}, including task formulation, multi-interest heterogeneous graph, and graph neural network.
\modelname is illustrated in Section~\ref{sec:model}. Experiment results, including model performance, ablation study, and parameter sensitivity, are shown in Section~\ref{sec:experiment}. 
Related works in the group-enhanced recommendation and multi-interest network embedding are introduced in Section~\ref{sec:related works}.
Finally, the conclusion and future research problems are given in Section~\ref{sec:conclusion}.

\section{Preliminary}\label{sec:pre}
In this section, we first formulate the group-enhanced personalized recommendation task, and then we illustrate the multi-interest heterogeneous graph proposed in this paper.

\subsection{Task Formulation}

We have a set of users $\mathcal{U} = \{u_1,u_2,...,u_{\left | \mathcal{U} \right|}\}$, a set of items $\mathcal{V} = \{v_1,v_2,...,v_{\left | \mathcal{V} \right|}\}$ and a set of groups $\mathcal{G} = \{g_1,g_2,...,g_{\left | \mathcal{G} \right|}\}$. 
The interactions of users, items, and groups form three graphs: user-item graph $G_{\mathcal{UV}}$, group-item graph $G_{\mathcal{GV}}$, and group-user graph $G_{\mathcal{GU}}$. 
We let $\textit{\textbf{R}}_u \in \mathbb{R^{\left| \mathcal{U} \right| \times \left| \mathcal{V} \right|}}$ and $\textit{\textbf{R}}_g \in \mathbb{R^{\left| \mathcal{G} \right| \times \left| \mathcal{V} \right|}}$ denote user-item interaction matrix and group-item interaction matrix respectively, where $r_{uv}$ (or $r_{gv}) = 1$ if user $u$ (or group $g)$ purchased item $v$, otherwise  $r_{uv}$ (or $r_{gv}) = 0$. 
Similarly, $\textit{\textbf{S}} \in \mathbb{R^{\left| \mathcal{G} \right| \times \left| \mathcal{U} \right|}}$ represents group-user interaction matrix, where $s_{gu} = 1$ if group $g$ contains user $u$, otherwise $s_{gu} = 0$. 
We define $U(g)$ to denote the set of users in group $g$.
The primary objective is to predict how likely a user $u$ would adopt item $v$ based on historical user-item interactions and the group information.

\subsection{Multi-interest Heterogeneous Graph}

To integrate user-item graph $G_{\mathcal{UV}}$, group-item graph $G_{\mathcal{GV}}$, and group-user graph $G_{\mathcal{GU}}$ into a unified graph representation, we propose the multi-interest heterogeneous graph. 
\modelname introduces a set of interests $\mathcal{I}=\{\mathcal{I}^1, \mathcal{I}^2, \cdots, \mathcal{I}^M\}$ where $M$ is the number of interests. $\mathcal{I}^n = \{i_{u_1}^n, \allowbreak i_{u_2}^n, \cdots, i_{u_{\left | \mathcal{U} \right |}}^n\}$ is the set of all users' $n$-th interest, $n \leq M$.

\begin{definition}
\textbf{Multi-interest Heterogeneous Graph}.
The multi-interest heterogeneous graph is defined as $G = (V, E)$ where $V = (\mathcal{U}, \mathcal{V}, \mathcal{G}, \mathcal{I})$ denotes the node set and $E = (E_{uv}$, $E_{ui}$, $E_{gi}$, $E_{gv})$ denotes the edge set. $\mathcal{U}$, $\mathcal{V}$, $\mathcal{G}$, and $\mathcal{I}$ represent the set of users, items, groups, and interests. $E_{uv}$, $E_{ui}$, $E_{gi}$ and $E_{gv}$ denote the user-item edges, user-interest edges, group-interest edges, and group-item edges, respectively.
\end{definition}

\begin{figure}[htbp]
    \centering
    \includegraphics[scale=0.35]{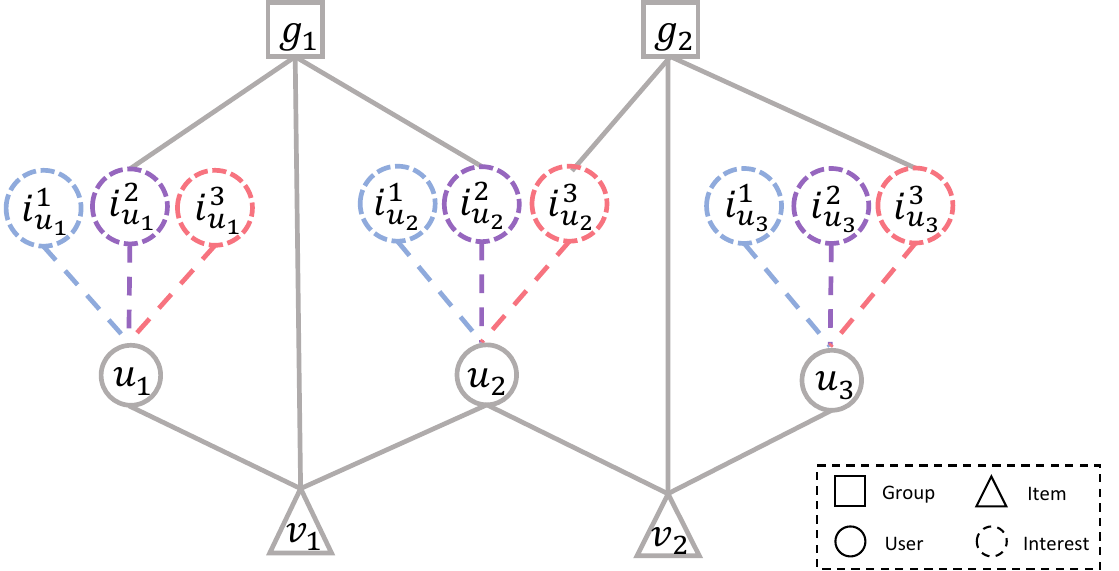}
    \caption{Multi-Interest Heterogeneous Graph. }
    \label{fig:hetegraph}
\end{figure}

The defined multi-interest heterogeneous graph is shown in Fig.~\ref{fig:hetegraph}. It can be split into four parts based on different types of edges. The first part is a user-item ($u$-$v$) bipartite graph representing user-item interaction.
An edge between $u$ and $v$ indicates the user $u$ has an interaction with the item $v$. 
The second part is a tree-structured user-interest ($u$-$i$) graph. Users' interests latently existed in the data, and \modelname disentangles the interests as the dashed interest nodes ($i$) in Fig.~\ref{fig:hetegraph}. Each user node is connected with a fixed number of interest nodes representing the user's interests. The third part is a group-interest ($g$-$i$) graph. It is also a tree-structured graph. Each group node has edges with one or more specific interest nodes of its group members. The last part is a group-item ($g$-$i$) graph that encodes group-item interactions.





\section{THE PROPOSED MODEL: \modelname}\label{sec:model}
\begin{figure*}[htbp]
    \centering
    \includegraphics[scale=0.35]{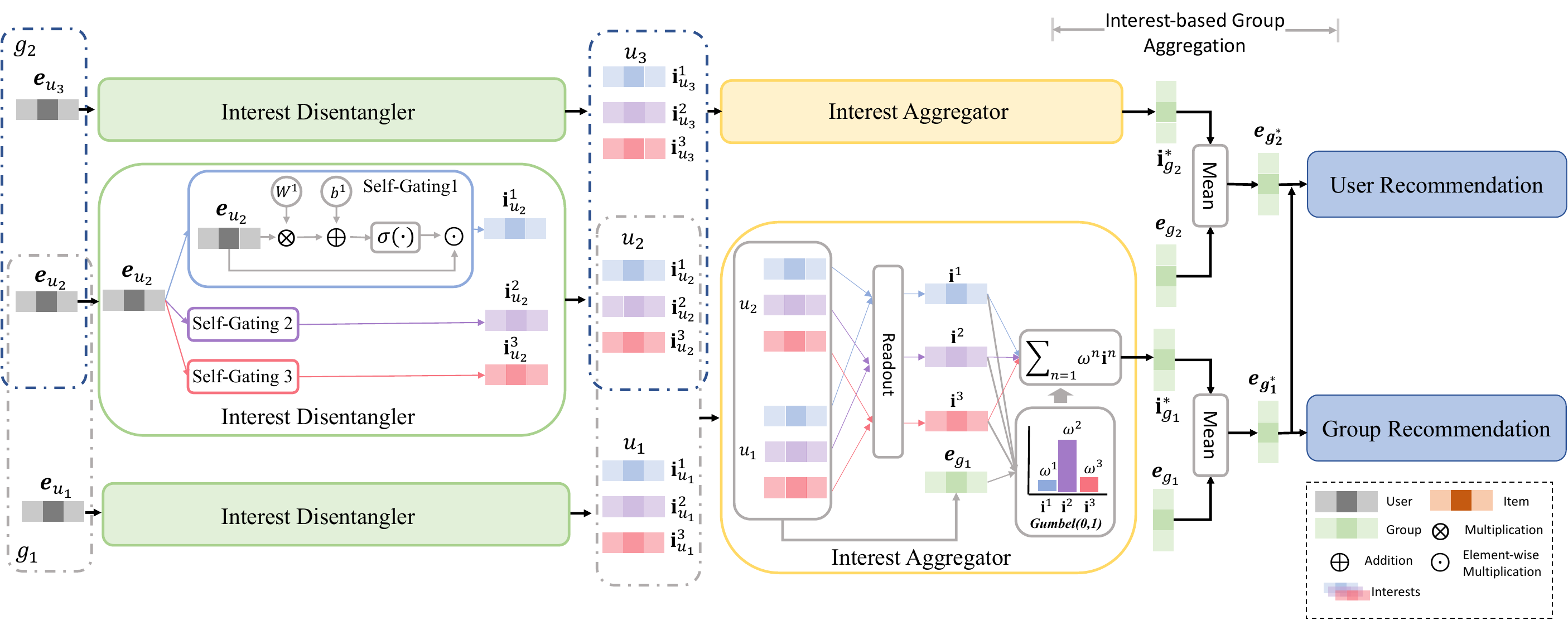}
    \caption{The framework of \modelname. 
    In this illustrated example, $u_2$ forms group $g_1$ with $u_1$ and group $g_2$ with $u_3$. 
    Based $g_1$ and $g_2$, \modelname predicts $u_2$'s rating score towards item $v_1$. \modelname consists of $4$ modules: 1) Interest disentangler that disentangles each user vector into interest vectors via self-gating. 
    2) Interest aggregator that selects and aggregates the relevant interests of the members within each group via gumbel sampling to obtain interest-based group representation. 
    3) Interest-based group aggregation that fuses $u_2$'s representation with the participated group representation. 4) Dual-trained rating predictor on both personalized recommendation and group recommendation.}

    \label{fig:mesh3}
\end{figure*}

The proposed \textbf{\modelname} is shown in Figure~\ref{fig:mesh3}, which consists of Interest Disentangler, Interest Aggregator, Interest-based Group Aggregation, and Dual-trained Rating Predictor.

\subsection{Interest Disentangler}

The core idea is that we assume users participate in different groups out of different interests in items, which is to solve the \textbf{multiple user interests} challenge. For example, in Figure~\ref{fig:mesh1}, Alice joins the "Competitive" group since she enjoys the competitive game CS:GO and joins the "Action" group because of her interest in the action game Hollow Night. Moreover, these interests should follow the user's overall representation. \modelname utilizes \textit{self-gating unit}~\cite{DBLP:conf/www/YuYLWH021} to disentangle user's interests from the user embedding. It computes a threshold on each dimension as a filter to extract the user's interests. 

Given a user node $u$ and its embedding $\mathbf{e}_u$, interest disentangler aims to disentangle $\mathbf{e}_u$ into $M$ interests $\mathbf{i}_u^1, \mathbf{i}_u^2,\cdots $ and 
$\mathbf{i}_u^M$. In previous works~\cite{MIND, Re4, ComiRec}, interests are generated randomly or learned from different interaction graphs. In this paper, we use multiple \textit{self-gating units} to disentangle user embedding into different interests. By passing $\mathbf{e}_u$ through these units and then each unit generates one interest representation for this user $u$, which is formulated as:
\begin{align}
    \mathbf{i}_{u}^{n} = f_\text{gate}^{n}(\mathbf{e}_u) = \mathbf{e}_u \odot \sigma(\mathbf{e}_u \mathbf{W}^n + \mathbf{b}^n),
\end{align}
where $\mathbf{i}_{u}^{n}$ denotes $u$'s $n$-th interest vector. $f_\text{gate}^{n}(\cdot)$ is the self-gating function that extracts $u$'s $n$-th interest. $\mathbf{W}^n \in \mathbb{R}^{d \times d}$, $\mathbf{b}^n \in \mathbb{R}^d$ are trainable parameters. $d$ is the embedding size. $\odot$ denotes the element-wise product and $\sigma$ is the sigmoid function. The left part of Fig.~\ref{fig:mesh3} illustrates the interest disentangler. 

\subsection{Interest Aggregator}

After obtaining the interests of each user, we next introduce the interest aggregator of each group $g$ to learn from \textbf{heterogeneous group members}. as shown in the middle part of Fig.~\ref{fig:mesh3}. We argue that a group is formed based on the user's specific interests, and \modelname utilizes an interest aggregator for each group to select and aggregate the relevant interests. Different from previous works~\cite{AGREE, GroupIM, MoSAN} that obtain the group representation by directly aggregating group members' overall embeddings, \modelname learns interest-based group representation by aggregating its members' selected interests. The probability of interest selection given memberships is calculated by the softmax formula as follows:
\begin{align}
    p(\delta(g) = n | U(g), g) 
    = \frac{\text{exp}(\mathbf{e}_g \cdot \text{Readout} (\{ \mathbf{i}_u^n \}_{u \in U(g)}))}{ {\textstyle \sum_{j=1}^{M}} \text{exp}(\mathbf{e}_g \cdot \text{Readout} (\{ \mathbf{i}_u^j \}_{u \in U(g)})) }.
    \label{eq:2}
\end{align}
To simplify the notation, we drop the subscript on group, where $\mathbf{e}_g$ represents group embedding for group $g$. $\cdot$ denotes the dot product. $\delta(g)$ is the currently selected interest for group node $g$. Readout function: $\mathbb{R}^{| \{ \mathbf{i}_u \}_{u \in U(g)} | \times d} \to \mathbb{R}^d$ is applied to aggregate group member's common interest information of all members within the group.

\subsubsection{Readout Function}

The readout function plays an important role in the message aggregating phase in GNNs. It summarizes an arbitrary number of embedding vectors into a fixed vector representation. Mean, Max, and Sum are the most common readout functions. 
We utilize Attention-based pooling as it is explainable that the interest of each user node contributes unequally to this group. 
Moreover, it also achieves the best performance in our experiments.
\begin{align}
    \gamma_{\mathbf{i}_u^n} = \frac{\text{exp}(\mathbf{W}_{\text{Att}} \mathbf{i}_u^n)}{\sum_{j \in U(g)} \text{exp}(\mathbf{W}_{\text{Att}} \mathbf{i}_j^n)}, \\
    \mathbf{i}^n = \text{Readout} (\{ \mathbf{i}_u^n \}_{u \in U(g)}) 
    = \sum_{j \in U(g)} \gamma_{\mathbf{i}_j^n} \mathbf{i}_j^n,
\end{align}
where $\mathbf{W}_{\text{Att}} \in \mathbb{R}^{1 \times d}$ is the trainable attention parameters. $\gamma_{\mathbf{i}_u^n}$ is calculated by the softmax formula and represents the learned weight of $\mathbf{i}_u$ in group $g$ with respect to the $n$-th interest. The readout function outputs the weighted sum of members' $n$-th interest vectors. $\mathbf{i} = (\mathbf{i}^1, \mathbf{i}^2, ..., \mathbf{i}^M) \in \mathbb{R}^{d \times M} $ denotes the aggregated interest embedding.

\subsubsection{Interest Selection via Gumbel Sampling}
The gumbel sampling method is used to determine the interest of the group formation. Gumbel-Softmax works as a differentiable method for categorical sampling via its reparameterization trick ~\cite{Gumbel}. We can obtain the Gumbel noise $\mathbf{G_n} \sim \text{Gumbel}(0,1)$ by:
\begin{align}
    G_n = -\text{log}(-\text{log}(\varepsilon _n)), \quad \varepsilon_n \sim \text{Uniform}(0,1).
\end{align}

Then, we sample from the following equation and draw the $M$-dimensional vector $\mathbf{\omega} = (\omega^1, \omega^2, ..., \omega^M) \in \mathbb{R}^{M}$.
\begin{align}
    \omega^n = \text{Softmax}((\log(\psi_n) + G_n) / \tau), \quad \psi_n = \textbf{e}_g \cdot \mathbf{i}^n,
\end{align}
where $\psi_n$ denotes the relevance score of the $n$-th interest under the distribution and $\tau$ is the temperature that approximate $\omega$ to a one-hot vector as $\tau$ goes to 0. Then, the Gumbel-Softmax $\omega^n$ is assigned as the probability of interest selection as $p(\delta(g) = n | \{ u \}_{u \in U(g)}, g)$ shown in Eq.~\ref{eq:2}. 
Finally, for group $g$, the learned interest-based vector $\mathbf{i}_g^{\ast} \in \mathbb{R}^{d}$ is calculated by multiplying $\omega$ and the aggregated interest vectors $\mathbf{i}$:
\begin{align}
    \mathbf{i}_g^{\ast} = \sum_{n=1}^{M} \omega^n \mathbf{i}^n .
\end{align}
Gumbel-Softmax has two versions: hard and soft. 1) Hard version: The output M-dimensional vector $\omega$ becomes a one-hot vector, which means only the most important interest will be kept. 2) Soft version: It is similar to the general softmax, but it assigns a larger weight to the most essential interest. We propose that users participate in a group for one specific interest or the combination of some interests, which corresponds to the hard Gumbel-Softmax and soft Gumbel-Softmax separately. Our experiments show that soft Gumbel-Softmax achieves better performance than the hard version.

\subsection{Interest-based Group Aggregation}
In this module, we achieve the \textbf{effective group
representation for personalized RS} by aggregating interest vectors. Given the aggregated interest vector $\mathbf{i}_g^{\ast}$ for group $g$, we collect the users' preferences from multiple interests with different weights. Then, \modelname updates group vector $\mathbf{g}^{\ast}$ as:
\begin{align}
    \mathbf{e}^{\ast}_g = (\mathbf{e}_g + \mathbf{i}_g^{\ast})/2, \label{eq:10}
\end{align}
The formation of a group is determined by its members, while the group also has implicit influences on each user. Therefore, we update the user's representation $\mathbf{e}_u$ by aggregating the embeddings of groups in which this user participates: 
\begin{align}
    \mathbf{e}_{\hat{u}} = (\mathbf{e}_u + \text{Pooling}(\{ \mathbf{e}^{\ast}_{g_l} \}_{g_l \in G(u)}))   / 2,
\end{align}
where $\mathbf{e}_{\hat{u}}$ is the updated user representation, $G(u)$ represents user $u$ participated group and Pooling($\cdot$) denotes the pooling operation for integrating the group information. We chose average pooling for its simplicity and outstanding performance in our experiments. 
\subsection{Dual-trained Rating Predictor}
\subsubsection{Personalized Recommendation Loss}


GNN-based models~\cite{LightGCN,DBLP:conf/sigir/YangLDMY21} have shown that graph neural network achieves better performance on recommendation tasks by aggregating the high-order user-item interactions. Here we utilize LGCN~\cite{LightGCN} on the user-item bipartite graph to model the relationship between the user and item as:
\begin{equation}
    \begin{gathered}
        \mathbf{e}_{\hat{u}}^{(k+1)} = \sum_{v \in \mathcal{N}_u} \frac{1}{\sqrt{\left | \mathcal{N}_u \right |} \sqrt{\left | \mathcal{N}_v \right |} } \mathbf{e}_{v}^{(k)}, \\
        \mathbf{e}_{v}^{(k+1)} = \sum_{u \in \mathcal{N}_v} \frac{1}{\sqrt{\left | \mathcal{N}_v \right |} \sqrt{\left | \mathcal{N}_u \right |} } \mathbf{e}_{\hat{u}}^{(k)},
    \end{gathered}
\end{equation}
where $\mathbf{e}_{\hat{u}}^{(k)}, \mathbf{e}_{v}^{(k)}$ denote the latent vectors of user $u$ and item $v$ after $k$ layers propagation. In the user-item bipartite graph, $\mathcal{N}_u$ denotes the set of items that interact with user $u$, and $\mathcal{N}_v$ denotes the set of users which interacts with item $v$. After $k$ layers propagation, we fuse these embeddings obtained at each layer into the final embedding for each user and item:
\begin{align}
    \begin{aligned}
        \mathbf{e}_u = \sum_{k=0}^{K} \mathbf{e}_{\hat{u}}^{(k)},
        \mathbf{e}_v = \sum_{k=0}^{K} \mathbf{e}_{\hat{v}}^{(k)}.
    \end{aligned}
\end{align}
\modelname predicts the preference score from $u$ to $v$ by the inner product of their final representations as:
\begin{align}
    r_{u,v} = \mathbf{e}_{u} \cdot \mathbf{e}_v.
\end{align}
We use the Bayesian Personalized Ranking (BPR) loss~\cite{BPR} to optimize the model for top-N recommendation as: 
\begin{align}
    \mathcal{L}_{bpr} = - \sum_{(u, v, j) \in \mathcal{S}_{user}} \text{log} \sigma(r_{u, v}- r_{u, j}), 
\end{align}
where the $\mathcal{S}_{user}$ is a set of triplets $(u, v, j)$ for users, $v$ is the interacted item for $u$ and $j$ is a negative sampled item.

\subsubsection{Group Recommendation Loss}
We further conduct the group recommendation task to illustrate the effectiveness of interest-based group representation. Similar to the personalized (user) recommendation task, BPR loss is adopted for this task: 
\begin{align}
\mathcal{L}_{group} = - \sum_{(g, v', j') \in \mathcal{S}_{group}} \text{log} \sigma(\mathbf{e}^{\ast}_g \cdot \mathbf{e}_{v'} - \mathbf{e}^{\ast}_g \cdot \mathbf{e}_{j'}),
\end{align}
where $\mathcal{S}_{group}$ is a set of triplets $(g, v', j')$. $v'$ is the item that group $g$ interacted with and $j'$ is a sampled negative item.

\subsubsection{Interest Regularization}
Moreover, one interest embedding captures one certain interest. For example, in Figure~\ref{fig:mesh1}, the Action group is formed by Alice, Aaron, and Tiffany because of their specific interest in action games. They also participate in other groups of other interests. We further add a regularization term to distinguish different interests:
\begin{align}
    &\text{Reg}_{asp} = \sum_{u=1}^{\left | \mathcal{U} \right|} \sum_{p=1}^{M-1} \sum_{q=p+1}^{M} m_{p,q}^u \cdot \text{Sim}(\mathbf{i}_u^p, \mathbf{i}_u^q),
\end{align}
where $\text{Sim}(\cdot)$ is the cosine distance that denotes the interest similarity between $p$-th and $q$-th interest of user $u$. Therefore, the value range of $\text{Sim}(\cdot)$ is [-1, 1], which is calculated as:
\begin{align}
    \text{Sim}(\mathbf{i}_u^p, \mathbf{i}_u^q) = \frac{\mathbf{i}_u^p \cdot \mathbf{i}_u^q}{\left \| \mathbf{i}_u^p \right \| \left \| \mathbf{i}_u^q \right \|}. \label{eq:15}
\end{align}
Regularizing all interest pairs is inefficient and calculation-intensive, and we only regularize on interest pairs with high similarity by a binary mask $m_{p,q}^u$ controlled by a threshold $t$:
\begin{align}
    &m_{p,q}^u = 
    \begin{cases}
    \quad 1, \quad \left | \text{Sim}(\mathbf{i}_u^p, \mathbf{i}_u^q) \right | \ge t
    \\
    \quad 0, \quad \text{otherwise}
    \end{cases},
    \label{eq:17} 
\end{align}
where larger $t$ allows more information can be shared between different interests, and smaller $t$ forces interests to be diverse. Then the final loss function is defined as:
\begin{align}
    \mathcal{L} = \eta_1 \mathcal{L}_{bpr} + (1 - \eta_1) \mathcal{L}_{group} + \eta_2 \text{Reg}_{asp} + \lambda {\left \| \Theta \right \|}_2^2.
\end{align}
It consists of BPR loss, group recommendation loss, interest regularization, and parameter regularization. $\eta_1$, $\eta_2$ and $\lambda$ are the weights for different losses.
\section{Experiments}\label{sec:experiment}
In this section, we conduct extensive experiments on \modelname to answer the following Research Questions~(RQs):
\begin{itemize}[leftmargin=*]
    \item \textbf{RQ1}: Does \modelname outperform existing methods in group enhanced recommendation?
    \item \textbf{RQ2}: Can \modelname achieve better performance on group recommendation tasks with interest-based group representation?
    \item \textbf{RQ3}: Are the components in \modelname necessary?
    \item \textbf{RQ4}: What is the impact of different experiment settings? 
 
\end{itemize}

\subsection{Experiment Settings}

\subsubsection{Datasets}

We conduct experiments on three publicly available datasets: MaFengWo~\cite{SoAGREE}, Douban~\cite{Douban}, and Steam~\cite{game-dataset}. The details of the datasets can be seen in Table~\ref{table1}. MaFengWo is a tourism website that provides a platform for customers to create or join a group trip. Douban, a well-known social media platform in China, allows users to rate and share their experiences with movies and books they watched. Steam is an online gaming platform on which game players can form interest groups to discuss games.
Since Steam and Douban datasets do not contain the interactions between group and item, we follow \cite{GGRM} to generate group item interactions based on the top 30 most frequently purchased among all members to run methods that need group-item interactions. 

\begin{table}
  \caption{Statistics of the Datasets}
  \centering
  \label{table1}
  \begin{tabular}{l c c c}
        \toprule
        \textbf{Dataset} & \textbf{MaFengWo} & \textbf{Douban} & \textbf{Steam} \\
        \hline
        \textbf{\#Users} & 5,275 & 12,925 & 375,257\\

        \textbf{\#Items} & 1,513 & 10,783 & 2,580\\

        \textbf{\#U-I interactions} & 39,767 & 553,766 & 12,181,681\\
        \textbf{Density} & 0.498\% & 0.397\% & 1.258\% \\
        \hline
        \textbf{\#Groups} & 995 & 2,753 & 849\\

        \textbf{\#G-I interactions}  & 3,595 & 82,590 & 25,470 \\

        \textbf{Avg. group size} & 9.63 &  203.46 & 547.10\\
        \bottomrule
  \end{tabular}
\end{table}

\begin{table*}[t]
    \caption{Overall comparison, the best and second-best results are in bold and underlined, respectively}
    \label{table2}
    \small
    \centering
    \setlength{\tabcolsep}{1.2mm}{
    \scalebox{0.825}{
    \begin{tabular}{l|l|ccccccccc}
         \hline
         Dataset & Metric & Popularity & NeuMF & MF & LightGCN & GREE & AGREE & GGRM & \modelname & Improv.\\
         \hline
         \multirow{4}{*}{MaFengWo}&R@5& 0.1832 \scriptsize $\pm$ 0.0000 & 0.2312 \scriptsize $\pm$ 0.0378 & 0.2603 \scriptsize $\pm$ 0.0161 & 0.3259 \scriptsize $\pm$ 0.0234  & 0.2915 \scriptsize $\pm$ 0.0098 & 0.2914 \scriptsize $\pm$ 0.0081  & \underline{0.3387 \scriptsize $\pm$ 0.0372} & \textbf{0.3694 \scriptsize $\pm$ 0.0016} & 9.06\%\\
         
         & R@10 & 0.3334 \scriptsize $\pm$ 0.0000 & 0.3925 \scriptsize $\pm$ 0.0591 &  0.4285 \scriptsize $\pm$ 0.0188 & 0.4996 \scriptsize $\pm$ 0.0364 & 0.4499 \scriptsize $\pm$ 0.0032 & 0.4477 \scriptsize $\pm$ 0.0043 & \underline{0.5211 \scriptsize $\pm$ 0.0340} & \textbf{0.5463 \scriptsize $\pm$ 0.0039} & 4.84\%\\
         
         & N@5 &0.1297 \scriptsize $\pm$ 0.0000 & 0.1780 \scriptsize $\pm$ 0.0332 & 0.2020 \scriptsize $\pm$ 0.0156 & 0.2601 \scriptsize $\pm$ 0.0217 & 0.2237 \scriptsize $\pm$ 0.0054 &  0.2230 \scriptsize $\pm$ 0.0040 & \underline{0.2684 \scriptsize $\pm$ 0.0295} & \textbf{0.3036 \scriptsize $\pm$ 0.0058} & 13.11\%\\
         
         & N@10 & 0.1863 \scriptsize $\pm$ 0.0000 & 0.2386 \scriptsize $\pm$ 0.0410 & 0.2648 \scriptsize $\pm$ 0.0164 & 0.3242 \scriptsize $\pm$ 0.0165 & 0.2821 \scriptsize $\pm$ 0.0029 & 0.2809 \scriptsize $\pm$ 0.0022 & \underline{0.3362 \scriptsize $\pm$ 0.0284} & \textbf{0.3684 \scriptsize $\pm$ 0.0056} & 9.58\%\\
         \hline
        \multirow{4}{*}{Douban} & R@5  & 0.0444 \scriptsize $\pm$ 0.0000 & 0.0413 \scriptsize $\pm$ 0.0034 & 0.0713 \scriptsize $\pm$ 0.0026 & 0.0726 \scriptsize $\pm$ 0.0027 & 0.0571 \scriptsize $\pm$ 0.0008 & 0.0571 \scriptsize $\pm$ 0.0005 & \underline{0.0720 \scriptsize $\pm$ 0.0020} & \textbf{0.0814 \scriptsize $\pm$ 0.0015} & 13.06\%\\
        
         & R@10 & 0.0762 \scriptsize $\pm$ 0.0000 & 0.0721 \scriptsize $\pm$ 0.0043 & 0.1117 \scriptsize $\pm$ 0.0026 & 0.1164 \scriptsize $\pm$ 0.0025 & 0.0916 \scriptsize $\pm$ 0.0011 & 0.0915 \scriptsize $\pm$ 0.0013 & \underline{0.1143 \scriptsize $\pm$ 0.0027} & \textbf{0.1249 \scriptsize $\pm$ 0.0015} & 9.27\%\\
         
         & N@5 & 0.0479 \scriptsize $\pm$ 0.0000 & 0.0479 \scriptsize $\pm$ 0.0035 & 0.0748 \scriptsize $\pm$ 0.0021 & 0.0846 \scriptsize $\pm$ 0.0024 & 0.0759 \scriptsize $\pm$ 0.0003 & 0.0763 \scriptsize $\pm$ 0.0003 & \underline{0.0812 \scriptsize $\pm$ 0.0049} & \textbf{0.0992 \scriptsize $\pm$ 0.0009} & 22.17\%\\
         
         & N@10 & 0.0574 \scriptsize $\pm$ 0.0000 & 0.0567 \scriptsize $\pm$ 0.0031 & 0.0867 \scriptsize $\pm$ 0.0020 & 0.0952 \scriptsize $\pm$ 0.0021 & 0.0814 \scriptsize $\pm$ 0.0004 & 0.0819 \scriptsize $\pm$ 0.0005 & \underline{0.0923 \scriptsize $\pm$ 0.0040} &  \textbf{0.1085 \scriptsize $\pm$ 0.0010} & 17.55\%\\
         \hline
        \multirow{4}{*}{Steam} & R@5 & 0.1154 \scriptsize $\pm$ 0.0000 & 0.0780 \scriptsize $\pm$ 0.0208 & 0.3825 \scriptsize $\pm$ 0.0089 & 0.4268 \scriptsize $\pm$ 0.0050 & 0.1795 \scriptsize $\pm$ 0.0109 & 0.1941 \scriptsize $\pm$ { \scriptsize 0.0078} & \underline{0.4255 \scriptsize $\pm$ 0.0027} & \textbf{0.4511 \scriptsize $\pm$ 0.0034} & 6.02\%\\
        
         & R@10  & 0.2444 \scriptsize $\pm$ 0.0000 & 0.1415 \scriptsize $\pm$ 0.0322 & 0.4975 \scriptsize $\pm$ 0.0092 & 0.5526 \scriptsize $\pm$ 0.0041 & 0.3157 \scriptsize $\pm$ 0.0248 & 0.3429 \scriptsize $\pm$ 0.0041 &\underline{0.5513 \scriptsize $\pm$ 0.0016} & \textbf{0.5727 \scriptsize $\pm$ 0.0024} & 3.88\%\\
         
         & N@5 & 0.1100 \scriptsize $\pm$ 0.0000 & 0.0680 \scriptsize $\pm$ 0.0148 & 0.3944 \scriptsize $\pm$ 0.0090 & 0.4490 \scriptsize $\pm$ 0.0050 & 0.1839 \scriptsize $\pm$ 0.0054 & 0.1927 \scriptsize $\pm$ 0.0044 & \underline{0.4466 \scriptsize $\pm$ 0.0026} & \textbf{0.4762 \scriptsize $\pm$ 0.0037} & 6.63\%\\
         
         & N@10 & 0.1594 \scriptsize $\pm$ 0.0000 & 0.0927 \scriptsize $\pm$ 0.0194 & 0.4334 \scriptsize $\pm$ 0.0092 & 0.4895 \scriptsize $\pm$ 0.0047 & 0.2297 \scriptsize $\pm$ 0.0111 & 0.2437 \scriptsize $\pm$ 0.0038 & \underline{0.4868 \scriptsize $\pm$ 0.0021} & \textbf{0.5136 \scriptsize $\pm$ 0.0032} & 5.51\%\\
         \hline
    \end{tabular}
    }}
\end{table*}

\subsubsection{Baselines}
To demonstrate the effectiveness of \modelname, we compare it with three groups of representative baselines. 1) Non-personalized RS (Popularity~\cite{DBLP:conf/recsys/CremonesiKT10}) that does not differentiate user's personalized preference. 2) Methods without using group information (NeuMF~\cite{NeuMF}, MF~\cite{koren2009matrix}, LightGCN~\cite{LightGCN}) that recommend items purely based on user-item interactions. 3) Group-enhanced personalized RS (AGREE~\cite{AGREE}, GREE~\cite{AGREE}, GGRM~\cite{GGRM}) that predicts user preference by modeling both user-item interactions and group information.

Previous multi-interest works are applied on sequential recommendation tasks~\cite{MIND, ComiRec,Re4,DIN, UMI,MGNM,PIMI,KEMI,SINE}.
We also test the multi-interest frameworks by transforming user-item interactions into sequences and comparing their proposed multi-interest extractor layer with our interest disentangler module. However, our experiments show those methods are specifically designed for the sequential recommendation and perform much worse without items' sequential patterns as in our experimented dataset. Thus, we do not include them as baselines.


    
    
    
    
        
    

\subsubsection{Evaluation Method}

Same as AGREE~\cite{AGREE}, for each user(group), we randomly split her(its) interactions into training (80\%), validation (10\%), and test (10\%) sets. 
We tune hyper-parameters on the validation set and test the model's performance on the test set. Rank-based evaluation metrics include NDCG@K and Recall@K with K ranges in \{5,10\}. We test each model 5 times with different random seeds to ensure the stability of the result.


\subsubsection{Hyper-parameter Settings}
We use Adam~\cite{Adam} as the optimizer in our model. Hyper-parameters are tuned based on grid search. For the common hyper-parameters learning rate, weight decay, and embedding size, we tune them within the ranges of \{0.1, 0.05, 0.01, 0.005, 0.001\}, \{1e-1, 1e-2, 1e-3, 1e-4, 1e-5, 1e-6\} and \{8, 16, 32, 64, 128, 256, 512\}, respectively. 
We tune the number of LGCN for LightGCN, GGRM, and our model from 1 to 6. 
The number of interests $M$, temperature $\tau$, and interest regularization $\eta$ are the particular hyper-parameters in our model. We tune the number of interests $M$ with the range of \{2, 3, 4, 5, 6, 7, 8\}, and tune temperature $\tau$ and interest regularization $\eta$ from 0.1 to 1 with interval 0.1. 

\subsection{Performance Comparison (RQ1)}
Experiment results on performance comparison are shown in Table~\ref{table2}. We can have the following observations.
\begin{itemize}[leftmargin=*]
    \item \modelname consistently outperforms other methods on the three datasets. On NDCG@5, \modelname has over $5\%$ improvement over the second-best method on the three datasets. The variance of \modelname is also similar to the best baseline. It shows the effectiveness of \modelname by enhancing RS with group information. It reveals that disentanglement in \modelname is effective in using group information.
        \item For other group recommendation models, GREE and AGREE perform worse than LightGCN. It shows joint learning on user-item and group-item bipartite graphs do not generalize to personalized recommendations well. \modelname and GGRM perform better than baselines without using group information (MF and LightGCN). It shows group information can be utilized to enhance the RS.
        \item GNN-based models such as LightGCN, GGRM, and IGRec exhibit notable superiority over other techniques. This underscores the GNN's proficiency in acquiring user and item representations by leveraging high-order connectivity.
\end{itemize}

\subsection{Group Recommendation Task (RQ2)}

\begin{table*}[t]
    \caption{Comparison of Top-K performance on three datasets with baselines in group recommendation problem. The best and second-best results are in bold and underlined, respectively}
    \label{table3}
    \small
    \centering
    \setlength{\tabcolsep}{1.2mm}{
    \scalebox{0.85}{
    \begin{tabular}{l|l|ccccccc}
         \hline
         Dataset & Metric & GREE & AGREE & GroupIM & SGGCF (ND) & SGGCF (ED) & \modelname & Improv.\\
         \hline
         \multirow{4}{*}{MaFengWo}&R@5& 
         0.0028 \scriptsize $\pm$ 0.0000 &
         0.0028 \scriptsize $\pm$ 0.0000 &
         0.3114 \scriptsize $\pm$ 0.0243 &
         0.3178 \scriptsize $\pm$ 0.0122 &
         \underline{0.3240 \scriptsize $\pm$ 0.0025} &
         \textbf{0.5477 \scriptsize $\pm$ 0.0340} &
          69.04\%\\
         
         & R@10 & 
         0.0055 \scriptsize $\pm$ 0.0051 &
         0.0033 \scriptsize $\pm$ 0.0005 &
         \underline{0.4173 \scriptsize $\pm$ 0.0180} &
         0.3607 \scriptsize $\pm$ 0.0044 &
         0.3658 \scriptsize $\pm$ 0.0075 &
         \textbf{0.6345 \scriptsize $\pm$ 0.0247} &
          73.46\%\\
         
         & N@5 & 
         0.0022 \scriptsize $\pm$ 0.0004 &
         0.0022 \scriptsize $\pm$ 0.0003 &
         0.1954 \scriptsize $\pm$ 0.0115 &
         0.2645 \scriptsize $\pm$ 0.0096 &
         \underline{0.2702 \scriptsize $\pm$ 0.0046} &
         \textbf{0.4456 \scriptsize $\pm$ 0.0256} &
         64.91\%\\
         
         & N@10 & 
         0.0031 \scriptsize $\pm$ 0.0017 &
         0.0024 \scriptsize $\pm$ 0.0004 &
         0.2324 \scriptsize $\pm$ 0.0085 &
         0.2795 \scriptsize $\pm$ 0.0076 &
         \underline{0.2849 \scriptsize $\pm$ 0.0055} &
         \textbf{0.4761 \scriptsize $\pm$ 0.0229} &
          67.11\%\\
         \hline
        \multirow{4}{*}{Douban} & R@5  & 
         0.5353 \scriptsize $\pm$ 0.0395 &
         0.5349 \scriptsize $\pm$ 0.0323 &
         - &
         \underline{0.5268 \scriptsize $\pm$ 0.0152} &
         \textbf{0.5350 \scriptsize $\pm$ 0.0131} &
         0.5143 \scriptsize $\pm$ 0.0055 &
         -3.87\%\\
        
         & R@10 & 
         0.6778 \scriptsize $\pm$ 0.0403 &
         0.6772 \scriptsize $\pm$ 0.0355 &
         - &
         0.6916 \scriptsize $\pm$ 0.0062 &
         \underline{0.6953 \scriptsize $\pm$ 0.0039} &
         \textbf{0.6987 \scriptsize $\pm$ 0.0035} &
          0.49\%\\
         
         & N@5 & 
         0.4941 \scriptsize $\pm$ 0.0278 &
         0.4945 \scriptsize $\pm$ 0.0225 &
         - &
         \underline{0.4731 \scriptsize $\pm$ 0.0120} &
         \textbf{0.4819 \scriptsize $\pm$ 0.0139} &
         0.4540 \scriptsize $\pm$ 0.0030 &
         -5.79\%\\
         
         & N@10 & 
         0.5593 \scriptsize $\pm$ 0.0286 &
         0.5597 \scriptsize $\pm$ 0.0245 &
         - &
         \underline{0.5489 \scriptsize $\pm$ 0.0081} &
         \textbf{0.5557 \scriptsize $\pm$ 0.0098} &
         0.5389 \scriptsize $\pm$ 0.0014 &
          -3.02\%\\
         \hline
        \multirow{4}{*}{Steam} & R@5 & 
         0.4174 \scriptsize $\pm$ 0.0485 &
         0.4663 \scriptsize $\pm$ 0.1249 &
         - &
         \underline{0.4935 \scriptsize $\pm$ 0.0127} &
         0.4930 \scriptsize $\pm$ 0.0123 &
         \textbf{0.5366 \scriptsize $\pm$ 0.0096} &
          8.73\%\\
        
         & R@10  & 
         0.5725 \scriptsize $\pm$ 0.0493 &
         0.6383 \scriptsize $\pm$ 0.0318 &
         - &
         \underline{0.7066 \scriptsize $\pm$ 0.0045} &
         0.7060 \scriptsize $\pm$ 0.0050 &
         \textbf{0.7611 \scriptsize $\pm$ 0.0029} &
          7.71\%\\
         
         & N@5 & 
         0.3838 \scriptsize $\pm$ 0.0439 &
         0.4273 \scriptsize $\pm$ 0.0074 &
         - &
         \underline{0.4355 \scriptsize $\pm$ 0.0140} &
         0.4352 \scriptsize $\pm$ 0.0138 &
         \textbf{0.4692 \scriptsize $\pm$ 0.0066} &
          7.74\%\\
         
         & N@10 & 
         0.4543 \scriptsize $\pm$ 0.0444 &
         0.5056 \scriptsize $\pm$ 0.0153 &
         - &
         \underline{0.5330 \scriptsize $\pm$ 0.0100} &
         0.5327 \scriptsize $\pm$ 0.0101 &
         \textbf{0.5721 \scriptsize $\pm$ 0.0044} &
          7.34\%\\
         \hline
    \end{tabular}
    }}
\end{table*}

Group recommendation focuses on providing personalized recommendations to a group of individuals rather than individual recommendations. 
In this section, we conduct experiments on three datasets to demonstrate that \modelname can achieve good performance on group recommendation tasks.


Except for AGREE and GREE models, two state-of-the-art group recommendation models: GroupIM~\cite{GroupIM} and SGGCF~\cite{SGGCF}, are chosen as our baselines. 
Experiment results are shown in Table~\ref{table3}. GroupIM's performance on Douban and Steam are excluded because it exceeds the memory capacity of our machine. ND and ED of SGGCF represent the node-dropout and edge-dropout augmentation methods, respectively. From the group recommendation experiments, we can have the following observations:

\begin{itemize}[leftmargin=*]
    \item On Recall and NDCG, \modelname consistently has a huge improvement over the second-best method on MaFengWo and Steam datasets. On NDCG@10, \modelname even outperforms more than $7\%$ over the second-best method. 
    It shows the effectiveness of interest-based group representation. 
    \item However, \modelname performs slightly worse than SGGCF on the Douban dataset. Compared with MaFengWo and Steam, where groups are formed spontaneously, Douban's group information is manually built from previous research~\cite{yin2019social}. Thus, the improvement on MaFengWo and Steam datasets aligns better with real-life situations.
\end{itemize}

\subsection{Ablation Study (RQ3)}
\begin{figure}
      \begin{center}
        \includegraphics[width=.2\textwidth]{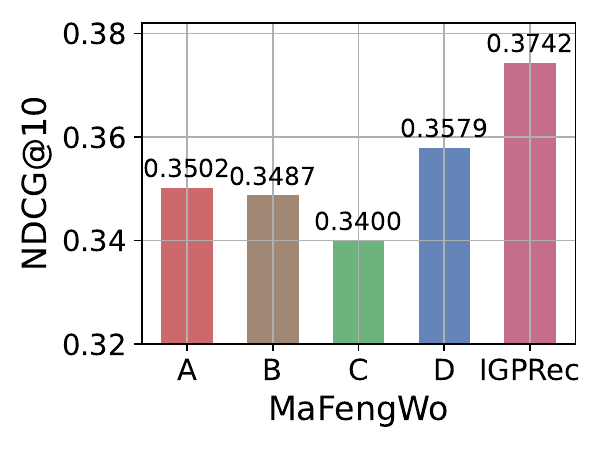}
        \includegraphics[width=.2\textwidth]{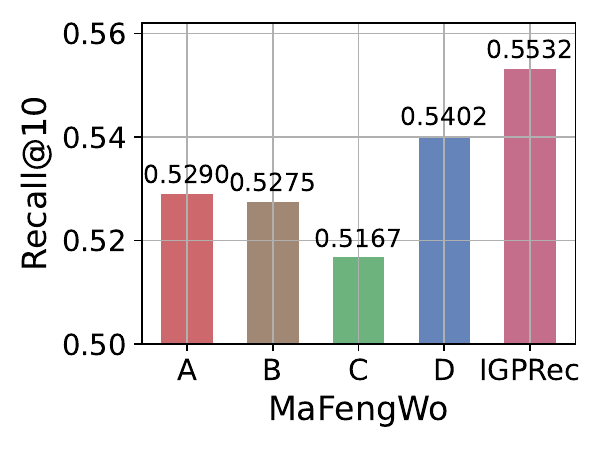}
        \includegraphics[width=.2\textwidth]{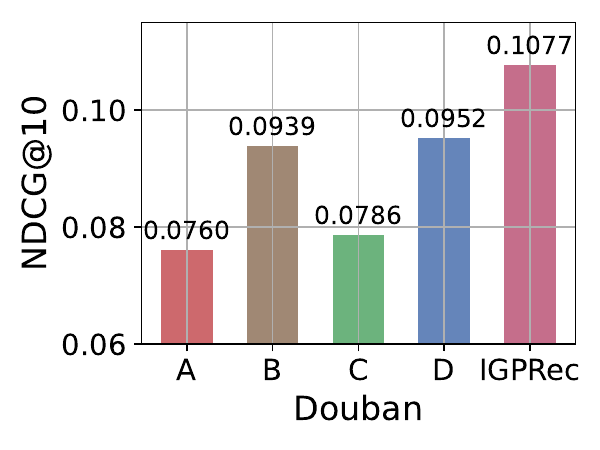}
        \includegraphics[width=.2\textwidth]{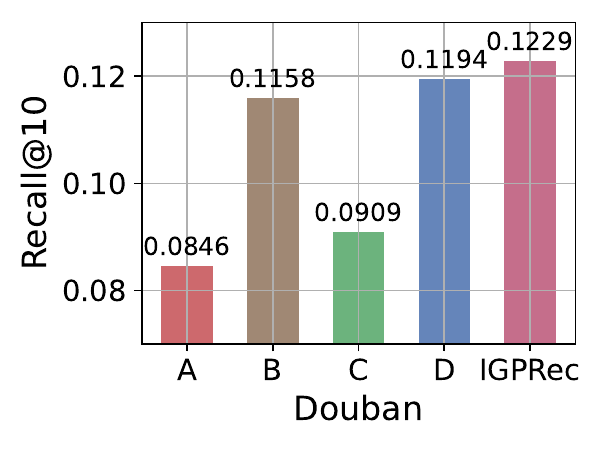}
        \includegraphics[width=.2\textwidth]{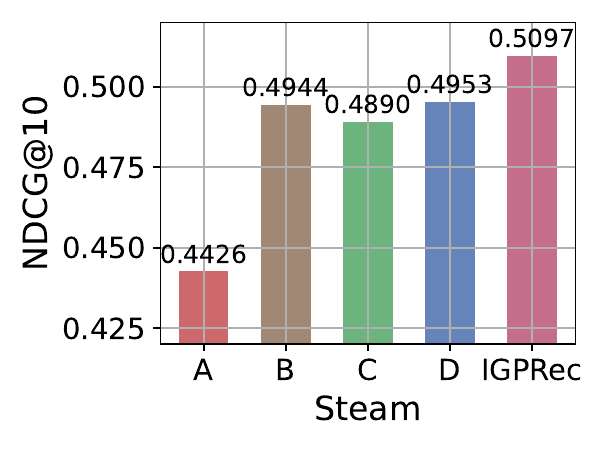}
        \includegraphics[width=.2\textwidth]{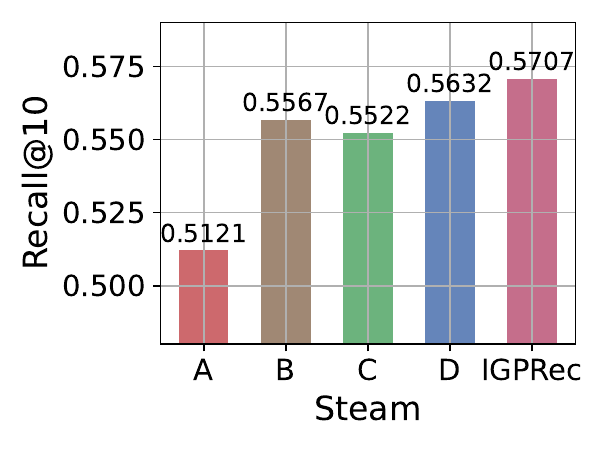}
      \end{center}
        \caption{Ablation study of \modelname.}
        \label{Ablation study}
\end{figure}
In this section, we conduct an ablation study to investigate the advantages of each component in \modelname. We consider 4 variants demonstrated in Fig.~\ref{Ablation study}. (1) Variant A removes the multi-interest module; (2) Variant B is built by replacing gumbel softmax by mean pooling on aggregating interests to the group. (3) Variant C removes the regularization of interests. (4) Variant D utilizes hard Gumbel-Softmax. 

We can have the following observations. (1) The regularization of interests has a critical impact on \modelname's performance. From the performance of variant C in each histogram, we find that removing the regularization of interests significantly influences the performance of \modelname. The model suffers from performance drop in all metrics, e.g., NDCG@10, 9.14\% in MaFengWo, 28.78\% in Douban, and 4.06\% in Steam. We aim to disentangle interests from users' embedding and let these interests be distinguishable. After removing the regularization of interests, it is hard to differentiate information among interests such that group embedding could not be representative corresponding to some specific interests. (2) The influence of multi-interest is highly related to datasets. Variant A learns group embeddings using mean pooling to directly aggregate members' embedding. The decrements of all metrics compared with \modelname are different in all datasets, which are 6.41\%, 29.43\%, and 13.16\% dropping ratio of NDCG@10 in MaFengWo, Douban, and Steam, respectively. This mainly depends on the property of datasets. Overall, it reveals that the multi-interests module is effective. 
(3) Soft Gumbel Softmax can further improve the performance. Variant B is the model that keeps all other modules but replaces Gumbel softmax with mean pooling to aggregate learned interests to the groups and variant D replaces soft Gumbel Softmax with the hard version. 
Among the four modules, soft Gumbel Softmax has the least improvement on \modelname.

\subsection{Hyper-parameter Analysis (RQ4)}

\begin{figure*}
     \centering
     \begin{subfigure}[b]{0.2\textwidth}
         \centering
         \includegraphics[width=\textwidth]{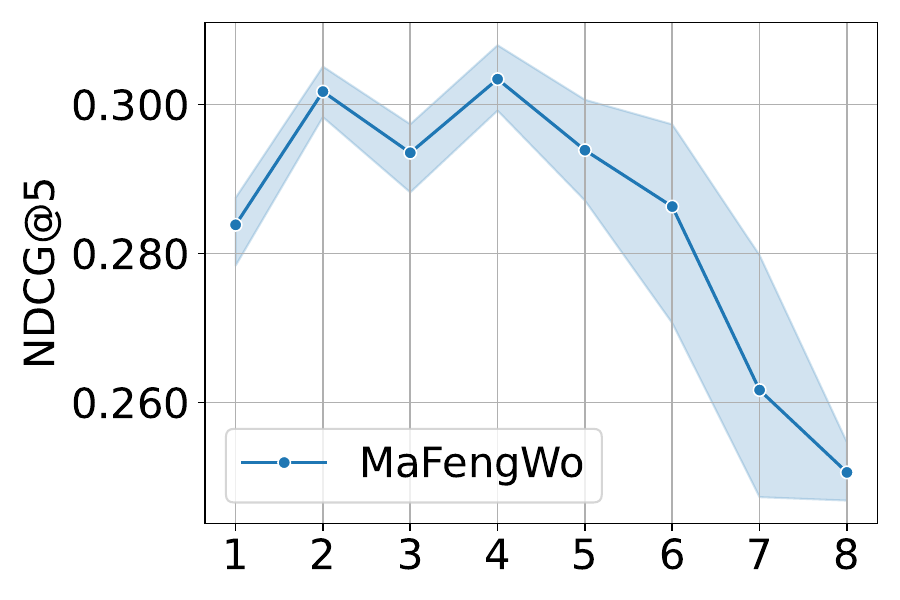}
         \caption{Number of Interests $M$}
         \label{MA Number of interests}
     \end{subfigure}
     \hfill
        \begin{subfigure}[b]{0.2\textwidth}
         \centering
         \includegraphics[width=\textwidth]{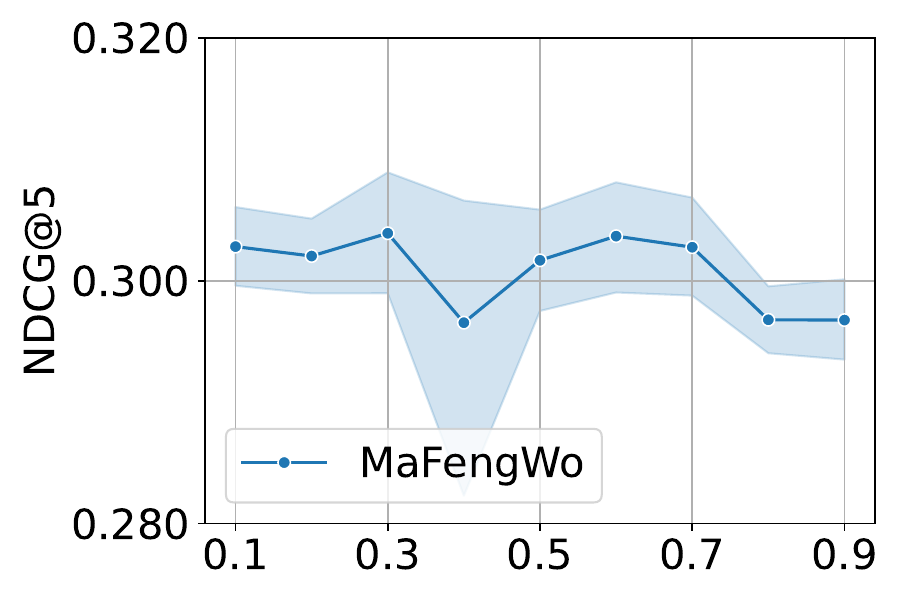}
         \caption{Interest Regularization $\eta$}
         \label{MA Interest Regularization}
     \end{subfigure}
      \hfill
    \begin{subfigure}[b]{0.2\textwidth}
         \centering
         \includegraphics[width=\textwidth]{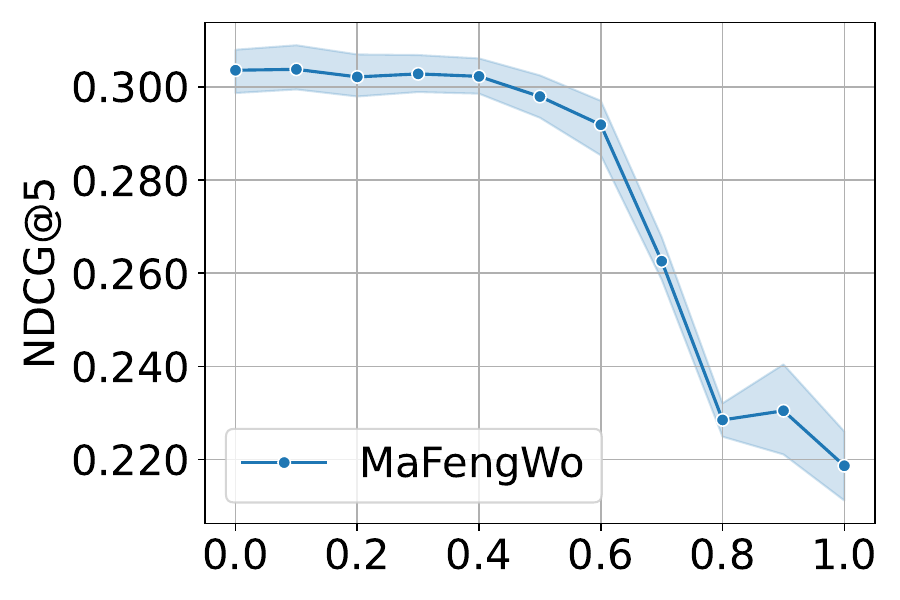}
         \caption{Similarity Threshold $t$}
         \label{MA Threshold}
     \end{subfigure}
      \hfill
    \begin{subfigure}[b]{0.2\textwidth}
         \centering
         \includegraphics[width=\textwidth]{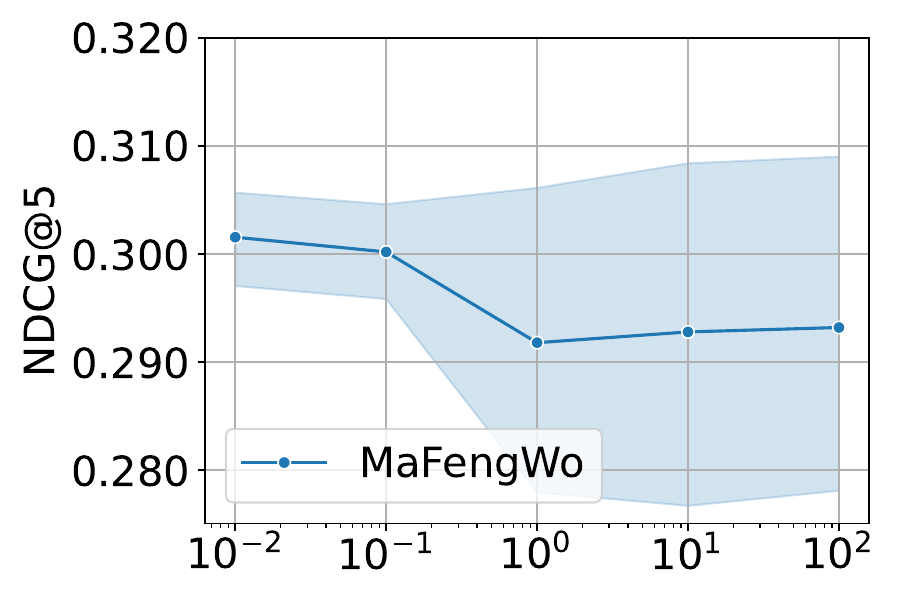}
         \caption{Temperature $\tau$}
         \label{MA temperature}
     \end{subfigure}

    \begin{subfigure}[b]{0.2\textwidth}
         \centering
         \includegraphics[width=\textwidth]{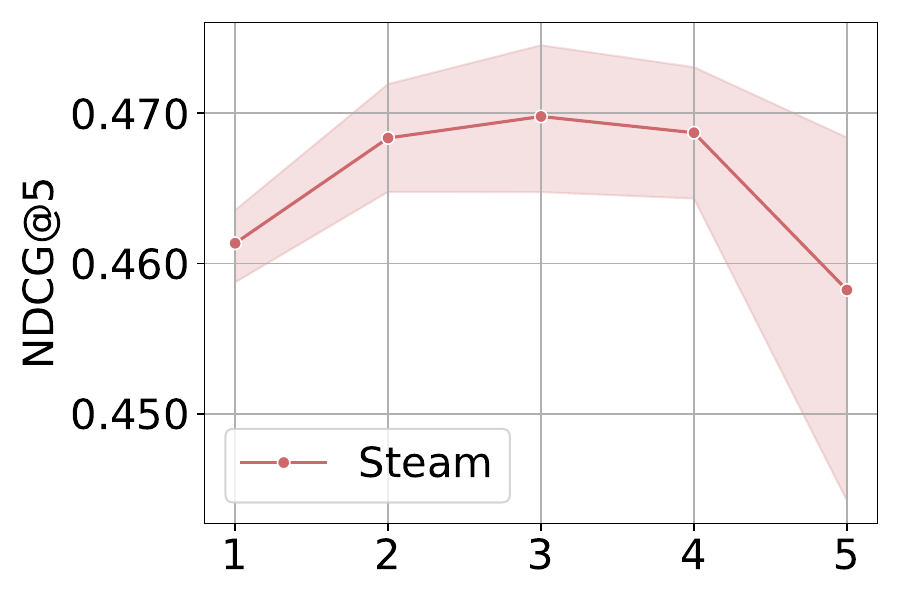}
         \caption{Number of Interests $M$}
         \label{Steam Number of interests}
     \end{subfigure}
     \hfill
     \begin{subfigure}[b]{0.2\textwidth}
         \centering
         \includegraphics[width=\textwidth]{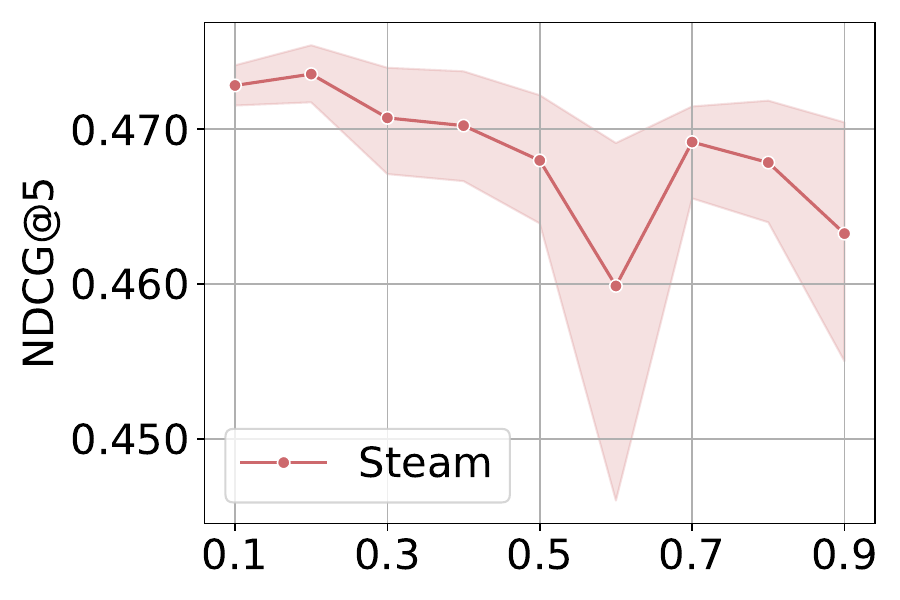}
         \caption{Interest Regularization $\eta$}
         \label{Steam Interest Regularization}
     \end{subfigure}
     \hfill
     \begin{subfigure}[b]{0.2\textwidth}
         \centering
         \includegraphics[width=\textwidth]{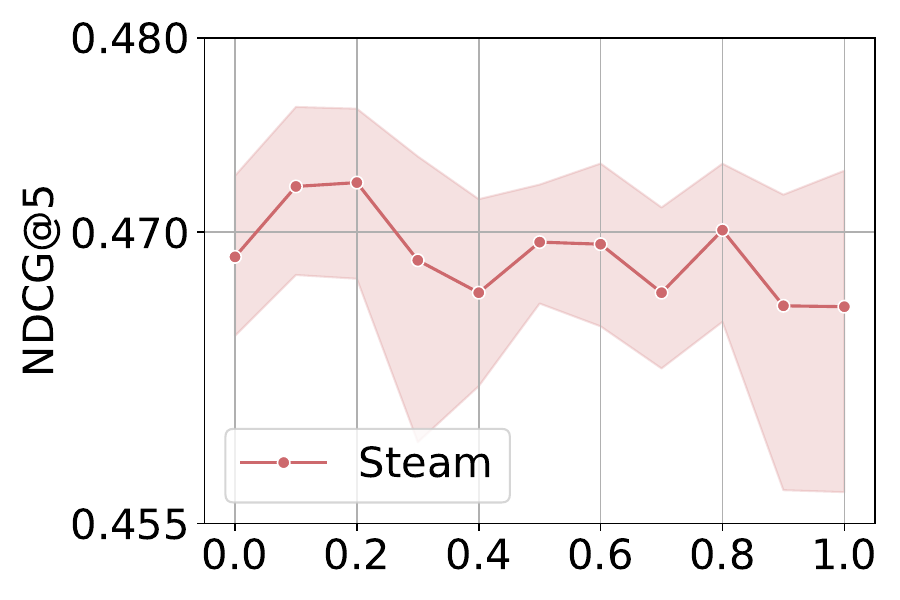}
         \caption{Similarity Threshold $t$}
         \label{Steam Threshold}
     \end{subfigure}
          \hfill
     \begin{subfigure}[b]{0.2\textwidth}
         \centering
         \includegraphics[width=\textwidth]{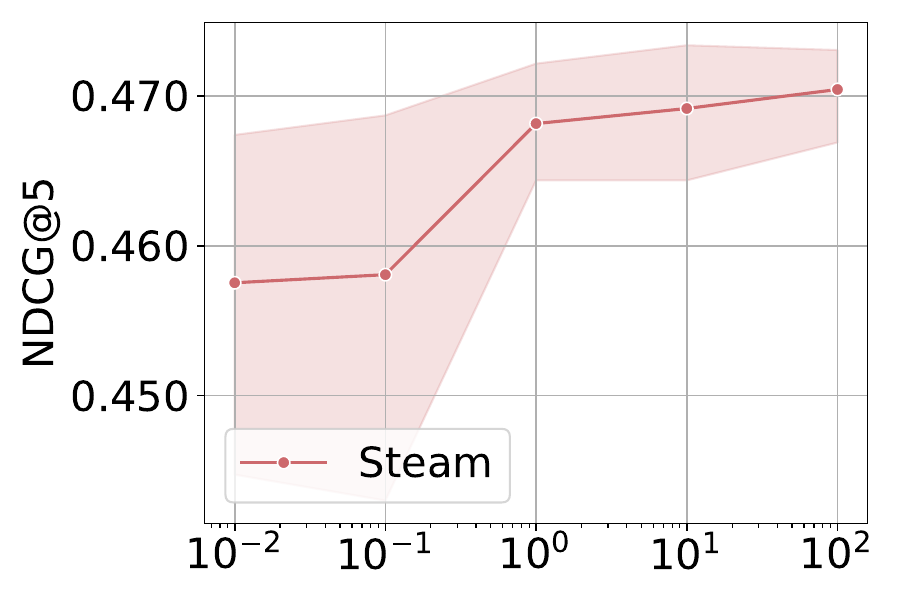}
         \caption{Temperature $\tau$}
         \label{Steam temperature}
     \end{subfigure}

        \caption{The influence of model hyper-parameters $M$, $\eta$, $t$, and $\tau$.}
        \label{Parameters}
\end{figure*}


\subsubsection{Hyper-parameter Analysis on $M$, $\eta$, $t$, and $\tau$.} The experiment results are shown in Fig.~\ref{Parameters}. 
The shadow space represents the variance. Due to the limited space, we only present experiment results on MaFengWo and Steam datasets. A similar trend is also observed in Douban. 
The number of interests $M$, interest regularization $\eta$, and threshold $t$ are three representative parameters in \modelname. For the number of interests from Fig.~\ref{MA Number of interests} and Fig.~\ref{Steam Number of interests}, we observe that the performance of NDCG@5 drops after achieving the peak point for each dataset. The range of $M$ for Steam dataset is from 1 to 5 due to the limited memory space.
Similarly, increasing the value of interest regularization $\eta$ will lead to better performance at the first few steps but fall rapidly when the value is too large, which is shown in Fig.~\ref{MA Interest Regularization} and Fig.~\ref{Steam Interest Regularization}. Fig.~\ref{MA Threshold} and Fig.~\ref{Steam Threshold} show the benefit of the threshold. Best performance is achieved when the threshold among interest embeddings is small, which leads to more diverse interests.
By contrast, the temperature parameter $\tau$ is the least important hyper-parameter to our model, illustrated in Fig.~\ref{MA temperature} and Fig.~\ref{Steam temperature}.

\begin{table}[t]
    \caption{Benefit of Self-Gating}
    \label{table4}
    \small
    \setlength{\tabcolsep}{1.2mm}{
    \scalebox{0.85}{
    \begin{tabular}{l|cccccc}
         \hline
         Method & R@5 & R@10 & N@5 & N@10 & \# Parameters\\
         \hline
        1 FC layer & 0.2721 & 0.4266 & 0.2131 & 0.2699 & $M \times (d+1) \times d$ \\
        2 FC layers & 0.2104 & 0.3418 & 0.1662 & 0.2144 & $2 \times M \times (d+1) \times d$ \\
         Interest embedding & \underline{0.3496} & \underline{0.5251} & \underline{0.2811} & \underline{0.3467} & $M \times \left |\mathcal{U}\right|\times d$ \\
         Self-Gating & \textbf{0.3694} & \textbf{0.5463} & \textbf{0.3036} & \textbf{0.3684} & $M \times (d+1) \times d$ \\
         \hline
         Improv. & 5.66\% & 4.04\% & 8.00\% & 6.26\% & \\
         \hline
    \end{tabular}
    }}
\end{table}

\subsubsection{Interest Generation Experiments} We compare self-gating with three methods, one and two Fully Connected (FC) layers and interest embedding. FC methods generate interests by linear layers, and interest embedding directly represents $M$ interests by $M$ embedding tables. Table~\ref{table4} shows that the self-gating outperforms the second best by about 5.35\% on Recall and 7.13\% on NDCG. Besides, self-gating and FC layers need much fewer parameters than interest embedding since $\left |\mathcal{U}\right| \gg (d+1)$. The self-gating mechanism achieves the best performance with fewer parameters, which effectively and efficiently generates interest from user embedding.

\begin{figure}
     \centering
     \begin{subfigure}[b]{0.22\textwidth}
         \centering
         \includegraphics[width=\textwidth]{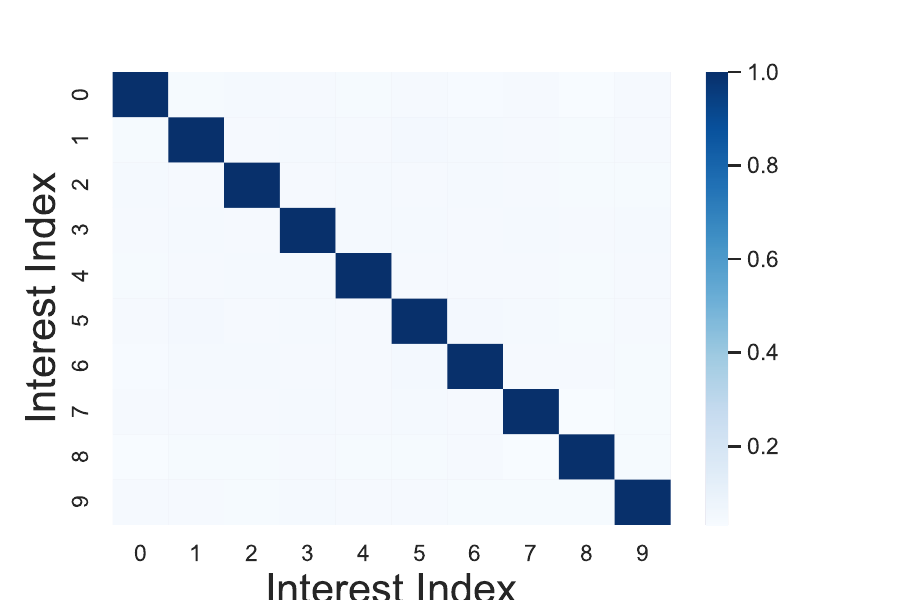}
         \caption{$t$ = 0.1}
         \label{Douban Threshold 0.1}
     \end{subfigure}
     \hfill
    \begin{subfigure}[b]{0.22\textwidth}
         \centering
         \includegraphics[width=\textwidth]{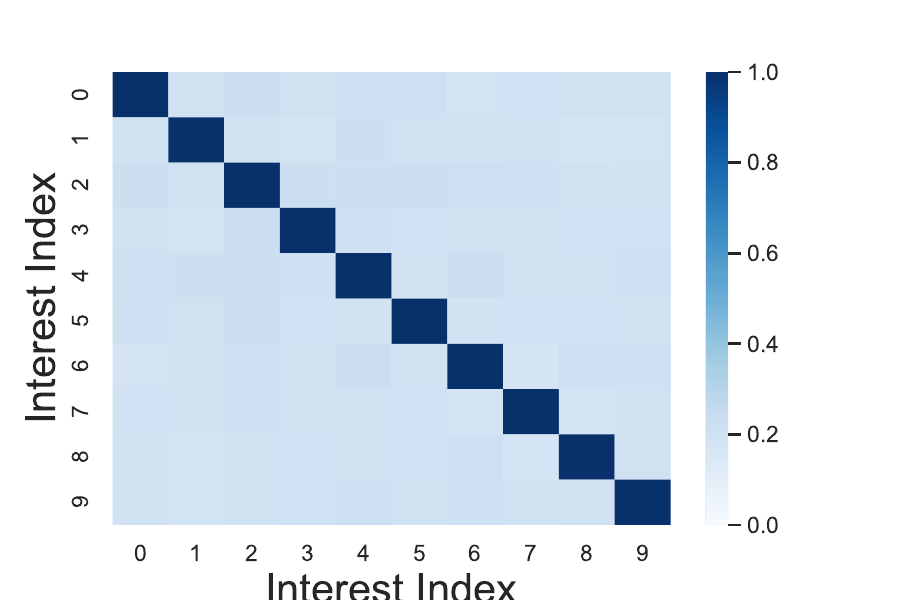}
         \caption{$t$ = 0.4}
         \label{Douban Threshold 0.4}
     \end{subfigure}
    \\
        \begin{subfigure}[b]{0.22\textwidth}
         \centering
         \includegraphics[width=\textwidth]{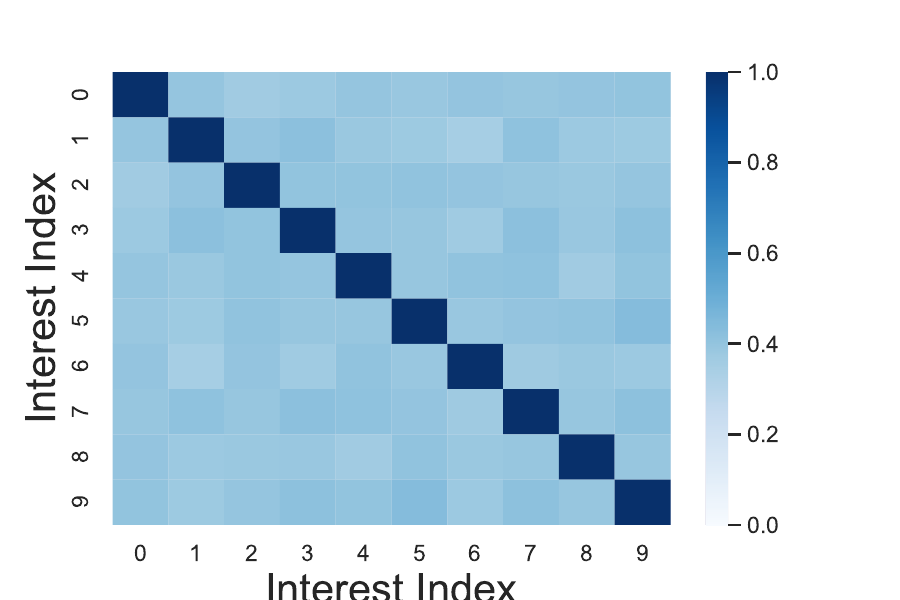}
         \caption{$t$ = 0.6}
         \label{Douban Threshold 0.6}
     \end{subfigure}
     \hfill
    \begin{subfigure}[b]{0.22\textwidth}
         \centering
         \includegraphics[width=\textwidth]{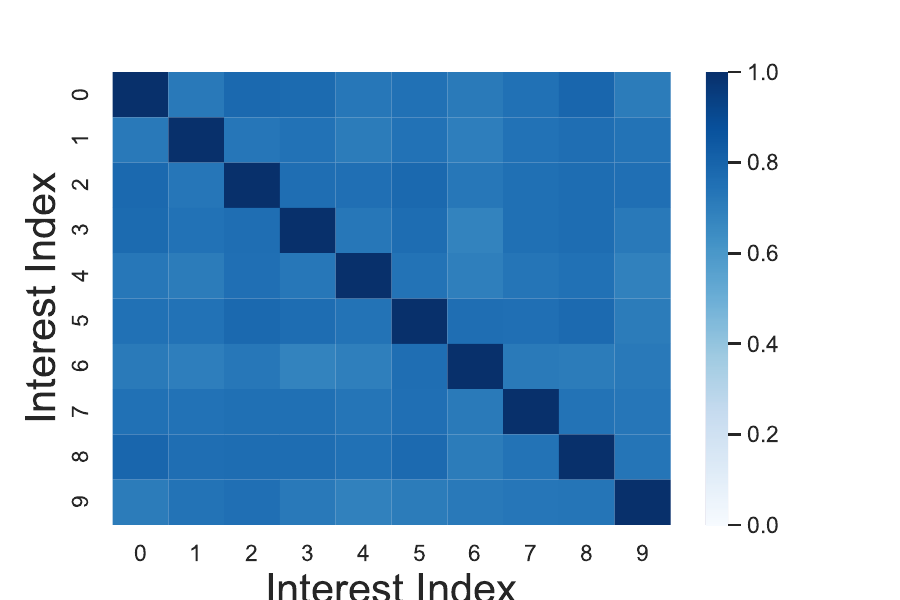}
         \caption{$t$ = 0.9}
         \label{Douban Threshold 0.9}
     \end{subfigure}
    \caption{The influence of threshold $t$ on Douban dataset}
        \label{Threshold}
\end{figure}

\subsubsection{The Influence of $t$ on interest similarity} The heatmaps in Fig.~\ref{Threshold} show the average pairwise cosine similarity between interest embeddings. According to Fig.~\ref{Douban Threshold 0.1}, each interest is distinguished compared with other interests when the threshold $t$ is 0.1. 
On the other hand, a larger $t$ allows interests to be similar (dark color) and share more common information with each other as shown in Fig.~\ref{Douban Threshold 0.4}, Fig.~\ref{Douban Threshold 0.6}, and Fig.~\ref{Douban Threshold 0.9}. 
 The best performance is achieved with a small threshold from Fig.~\ref{MA Threshold} and Fig~\ref{Steam Threshold}. It shows smaller $t$ helps achieve better performance by distinguishing the representation of the interests.

\subsubsection{The effect of $\eta_1$ on personalized recommendation and group recommendation}

\begin{figure}
     \centering
     \begin{subfigure}[b]{0.22\textwidth}
         \centering
         \includegraphics[width=\textwidth]{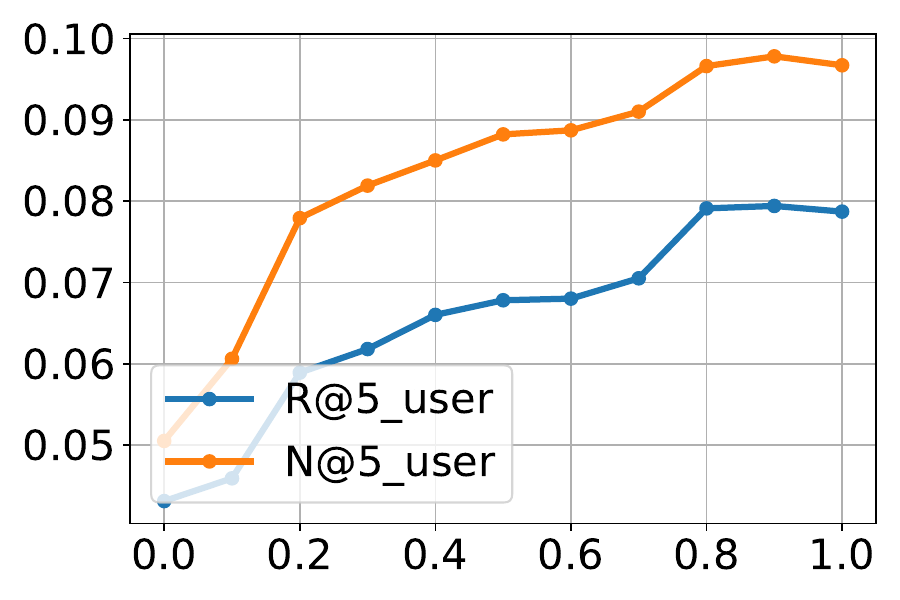}
         \caption{User Recommendation}
         \label{User Recommendation}
     \end{subfigure}
     \hfill
    \begin{subfigure}[b]{0.22\textwidth}
         \centering
         \includegraphics[width=\textwidth]{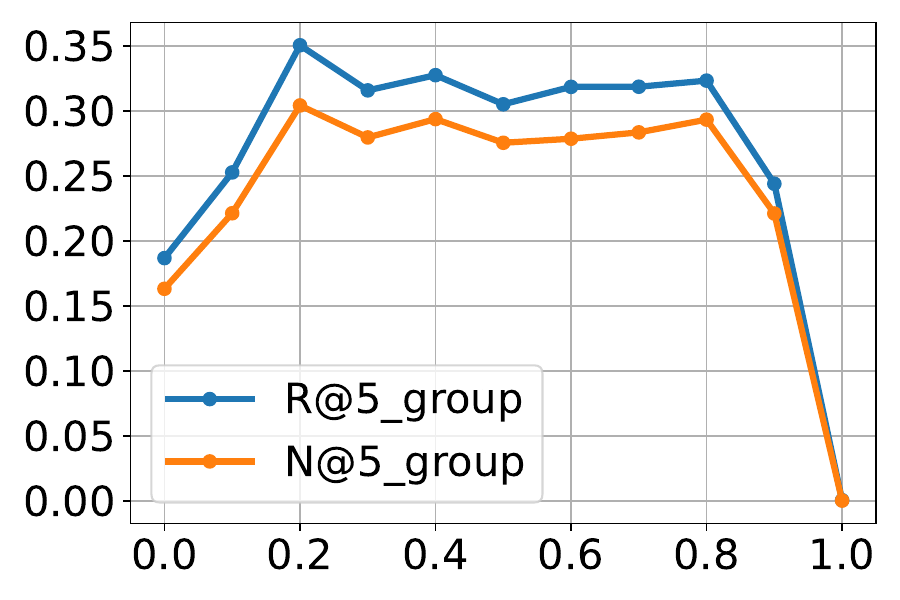}
         \caption{Group Recommendation}
         \label{Group Recommendation}
     \end{subfigure}
     \caption{The effect of $\mathcal{L}_{bpr}$ weight $\eta_1$ on Douban dataset}
        \label{eta_1}
\end{figure}

The line maps in Fig.~\ref{eta_1} show the changing trend of user recommendation and group recommendation performances along with the increment of $\eta_1$. 
According to Fig.~\ref{User Recommendation}, the performance on user recommendation increases steadily and achieves best when $\eta_1 = 0.9$.
For Fig.~\ref{Group Recommendation}, the best performance on group recommendation is achieved as $\eta_1 = 0.2$, which indicates the user recommendation task could promote the group recommendation. Then, the results decrease when increasing $\eta_1$ because the model focuses primarily on user recommendation tasks.

\section{Related Works}\label{sec:related works}
\subsection{Group Enhanced Recommendation}

RS that utilize group information can be classified into two categories: group-enhanced personalized RS and group-oriented RS.
 Group-enhanced personalized RS aims to recommend items to users by utilizing users' group participation as side information. To differentiate group members' contributions by attention mechanism, AGREE~\cite{AGREE} learns the dynamic aggregation from group members to infer the group’s decision on the item. Yuan et al.~\cite{yuan2009augmenting} consider users' community memberships and friendship connections as social connections and get improvements on the overall performance of recommendations via the fusion of social connections and collaborative filtering. Lee et al.~\cite{DBLP:journals/ipm/LeeB17} explore the feasibility and value of community-based sociality for a personalized recommendation. In order to capture preference dependency relationships between the target user and other group members, GGRM~\cite{GGRM} applies a graph neural network on three bipartite graphs (users-items, users-groups, and groups-items) to learn and concatenate user and item embedding for predicting.

Group-oriented RS aims to learn group representation and recommend items to a group of users. Previous works learn group representation by aggregating the preferences of its members with various predefined aggregation approaches, such as average~\cite{DBLP:conf/recsys/BaltrunasMR10}, least misery~\cite{DBLP:journals/pvldb/Amer-YahiaRCDY09} and maximum satisfaction~\cite{DBLP:series/sci/BorattoC11}. The predefined approaches cannot capture the complicated dynamic process of group RS. Recent works learn to differentiate group members' contributions by attention mechanism. MoSAN~\cite{MoSAN} models the user's preference with respect to all other members in the same group by sub-attention module. GroupIM~\cite{GroupIM} maximizes the mutual information between the user representation and their group representations which are aggregated from its members' preferences via the attention mechanism. Further, SGGCF~\cite{SGGCF} constructs a unified user-centered graph and implements group recommendation with self-supervised learning.

 Previous works directly aggregate the group members' embedding into group representation, which assumes users who participate in the same group share totally the same interests. They fail to consider that users join different groups out of different interests in real life, and a single group can only partially reflect the user's interests. Following this idea, \modelname disentangles the user's embedding into different interests and learns interest-based group representation.

\subsection{Multi-interest Recommendation}
The majority of RS models represent a single user by one vector with a fixed length. 
However, such a user representation might be less effective in capturing users' diverse interests.
Recently, learning multiple representations/interests of each user has been attracting much research attention~\cite{MIND, ComiRec,Re4,DIN, UMI,MGNM,PIMI,KEMI,SINE}. 
MIND~\cite{MIND} designs a multi-interest extractor layer with Behavior-to-Interest (B2I) dynamic routing for clustering users' behavior into interest representations and then chooses the interest associated with the specific item through a label-aware attention layer.
ComiRec~\cite{ComiRec} provides two methods to construct a multi-interest extraction module: dynamic routing method and self-attentive method. Additionally, it retrieves top-k items by a greed inference algorithm and introduces a controllable factor to trade off accuracy and diversity of recommendations.
Zhang et al.~\cite{Re4} devise a Re4 framework consisting of Re-contrast, Re-attend, and Re-construct. These three modules address the following problems: the distinction among interest representations, the consistency between the forward attention weights and the interest-item correlation, and the semantic reflection from interest on representative items. 

Taking full advantage of group information can help infer users' interest in items. Motivated by this, we introduce \modelname that fills this research blank and propose to learn interest-based group representations.

\section{Conclusion and Future Work}\label{sec:conclusion}
In this paper, we study how to enhance recommender systems with users' group information. 
Compared with previous works, we argue that users join different groups out of different reasons, and one group can only reveal one or more interests of group members.
We make the first attempt to disentangle users' interest in items via users' group participation and propose \modelname that disentangles users' embedding into interests and learns interest-based group representation. Experiment results show we significantly outperform the SOTA group enhanced personalized RS and group recommendation RS.

\section{Acknowledge}
This work is supported in part by NSF under grant III-2106758.

\bibliographystyle{IEEEtran}
\balance
\bibliography{ref_google}

\end{document}